\documentclass[copyright,creativecommons]{eptcs}
\providecommand{\event}{F-IDE 2019} 
\usepackage{breakurl}             
\usepackage{underscore}           

\usepackage{graphicx}
\usepackage[expansion = true, protrusion = true]{microtype}
\usepackage[utf8]{inputenc}
\usepackage[T1]{fontenc}
\usepackage{amssymb}
\usepackage[usenames,dvipsnames, table]{xcolor}
\usepackage{listings}
\usepackage{alloy}
\usepackage{hyperref}
\usepackage{xcolor}

\title{Simulation under Arbitrary Temporal Logic Constraints}
\author{Julien Brunel \qquad\qquad David Chemouil
\institute{ONERA DTIS and Université fédérale de Toulouse, France}
\and
Alcino Cunha \qquad\qquad Nuno Macedo
\institute{INESC TEC and Universidade do Minho, Portugal}
}
\def\titlerunning{Simulation under Arbitrary Temporal Logic Constraints}
\def\authorrunning{J. Brunel, D. Chemouil, A. Cunha, \& N. Macedo}
\begin{document}
\maketitle

\begin{abstract}
  Most model checkers provide a useful simulation mode, that allows
  users to explore the set of possible behaviours by interactively
  picking at each state which event to execute next. Traditionally
  this simulation mode cannot take into consideration additional
  temporal logic constraints, such as arbitrary fairness restrictions,
  substantially reducing its usability for debugging the modelled
  system behaviour. Similarly, when a specification is false, even if
  all its counter-examples combined also form a set of behaviours,
  most model checkers only present one of them to the user, providing
  little or no mechanism to explore alternatives. In this paper, we
  present a simple on-the-fly verification technique to allow the user
  to explore the behaviours that satisfy an arbitrary temporal logic
  specification, with an interactive process akin to simulation. This
  technique enables a unified interface for simulating the modelled
  system and exploring its counter-examples. The technique is
  formalised in the framework of state/event linear temporal logic and
  a proof of concept was implemented in an event-based variant of the
  Electrum framework.
\end{abstract}

\section{Introduction}

Model checking is one of the most successful techniques for analysing
systems, largely due to the ability to automatically verify whether a
temporal logic specification holds in a model of a system.  Model
validation and debugging is essential when analysing a system, and
most model checkers provide a simulation mode where the user can
explore alternative system traces by choosing how to proceed with the
exploration. With most tools, it is possible to choose one of the
possible successor states randomly. Additionally, in order to provide
a finer control and speed up the debugging process, many tools also
allow the user to interactively pick which event to execute next (if
the modelling language has some notion of event/action) and/or one of
the next possible states (to support the exploration of
non-deterministic events, both features must be provided). These
simulation modes are quite intuitive and can even be used by problem
domain experts unfamiliar with model checking to help validate the
model.

Unfortunately this simulation mode only takes into account the system model, traditionally specified by some sort of transition system or a set of events. However, in some situations it would be extremely helpful to perform such simulation under additional constraints, for example to assess the impact of imposing arbitrary fairness constraints. Such constraints reduce the set of valid behaviours and simulation could help the user validate and better understand their impact (which is not always trivial to infer). 
Similarly, when model checking a given temporal logic property it could be very useful to explore the set of behaviours that falsify it (its set of counter-examples) with a similar simulation technique.  Currently 
most model checkers display a single counter-example when a property is false. As a consequence, the user often
inspects the (lone) counter-example to locate the possible source of the
problem, changes the model (or specification) to address it, only for the
model checker to reveal a different counter-example to the same property. The
ability to explore distinct counter-examples at once could allow the user to
identify a more general fix, thus tightening the check / analyse / fix loop
and making the overall model checking process more efficient.

In this paper we propose a simulation technique that explores the set
of the behaviours that satisfy (or falsify) an arbitrary temporal
logic specification. At any point the user can
focus on a particular state of a trace, see which alternative events
enable the same trace prefix to be extended into a complete valid
behaviour (another infinite trace satisfying the property), and follow
any of those to proceed with the exploration. While traditional
simulation is rather easy to implement efficiently for any model
resembling a transition system, it is unclear how to do so when
additional constraints are imposed. This paper explores the viability
of a rather naïve on-the-fly technique: when a state is focused,
multiple queries to the model checker are run in the background to
determine which events can be further explored, while still preserving
the same trace prefix. To tame the complexity in models with many
events (or parametrised ones), type categorisation is supported: the
user first focuses on a specific type and only then iterates over the
different events of that type.

This paper is structured as follows. In the next section we very briefly discuss some alternative techniques to explore the set of behaviours that satisfy (or falsify) a given property. In Section \ref{sec:formalisation} we formalise our
proposal in the general setting of event/state linear temporal logic.
Section~\ref{sec:implementation} presents a prototype implementation of the
proposed technique in the Electrum Analyzer~\cite{electrumanalyzer}, the model
checker for the Electrum language~\cite{electrum}, an extension of Alloy~\cite{alloy} with
linear time temporal logic. The goal of this prototype is mainly to show the
viability of the approach, namely in terms of user-experience and efficiency.
Section~\ref{sec:conclusion} wraps-up the paper and presents some ideas for
future work.

\section{Related work}
\label{sec:related-work}

Some techniques have been proposed to explore of the set of behaviours that satisfy (or falsify) a given property. The simplest ones just provide
iteration over such set, by independently displaying one
trace at a time. This can be achieved by changing an explicit model checking
engine to resume search after finding one counter-example trace,
or, in the case of a SAT-based symbolic bounded model checker, by incrementally
adding new clauses that exclude exactly the previous trace, as implemented in
the Electrum Analyzer~\cite{electrumanalyzer} developed by the authors. The problem is
that this frequently keeps yielding traces that are just slight
variations of each other and, since the full set of behaviours is
usually too big to be enumerated, finding interesting variations may prove
infeasible. To alleviate this problem, for specific modelling languages it is
possible to define reasonable equivalence classes on traces (e.g., traces that
follow the same control-flow path are deemed equivalent), and implement
iteration by restarting the model checker with a modified property that
conjoins the original one with a formula excluding all traces in the class of
the previous counter-example~\cite{efsm,simulink}.

Problem domain expertise, namely some kind of user input, could lead to more
effective exploration. While in the above techniques user interaction is limited to
just asking for the next trace,
in \cite{marsha}, by running
multiple queries to the model checker, a proof tree of a CTL property is
inferred to ``explain'' a counter-example trace, with which the user can
interact to ask for new counter-examples. Possible interactions include asking
for alternative proofs (e.g., in a disjunction node), or guiding the search to
explore different parts of the model (e.g., in $\mathtt{EX}\
\phi$ nodes, by choosing the next $\phi$-satisfying state). However,
this approach requires substantial knowledge of the
underlying proof system for CTL and it is not clear how it can be generalised
to support LTL and fairness constraints.

\section{Formalisation}
\label{sec:formalisation}

\newcommand{\extends}{\prec}
\newcommand{\modifies}{{\cal F}}
\newcommand{\labels}{{\cal L}}
\newcommand{\events}{{\cal E}}
\newcommand{\after}{\mathop{\mathtt{X}}}
\newcommand{\always}{\mathop{\mathtt{G}}}
\newcommand{\eventually}{\mathop{\mathtt{F}}}
\newcommand{\until}{\mathop{\mathtt{U}}}
\newcommand{\forward}{\mathop{\triangleright\!\triangleright}}
\newcommand{\backward}{\mathop{\triangleleft\!\triangleleft}}
\newcommand{\nextstate}{\triangleright}
\newcommand{\nextevent}{\mathop{\blacktriangleright}}
\newcommand{\newevent}{\vartriangle}
\newcommand{\encode}[1]{[#1]}
\newcommand{\types}{\Upsilon}
\newcommand{\typeof}{{\cal T}}
\newcommand{\Nat}{\mathbb{N}}
\newcommand{\twodots}{\mathinner{\ldotp \ldotp}}

Most systems incorporate both the notion of states and events.
\emph{State/event linear temporal logic} (SE-LTL) was proposed to allow a more
concise and intuitive specification in these cases~\cite{stateevent}. The
semantics of a formula in this logic is defined over a \emph{labelled Kripke
structure} (LKS), a tuple $(S,I,P,\labels,T,\Sigma,\events)$ where $S$ is a
finite set of states, $I \subseteq S$ the set of initial states, $P$ a
finite set of atomic propositions, $\labels : S \to 2^P$ a state labelling
function, $T \subseteq S \times S$ a transition relation, $\Sigma$ a
finite set of events, and $\events : T \to 2^\Sigma \setminus \{\emptyset\}$
a transition labelling function. The transition relation is assumed to be
total, so every state has at least one successor.
To enable a more efficient exploration, events are categorized with a function
$\typeof : \Sigma \to \types$ that assigns a type to each event.
This categorization is natural in many models, namely those with parametrised
events. A \emph{path} $\pi = \langle s_0,a_0,s_1,a_1,\ldots \rangle$ of such a
\emph{typed LKS}
is an alternating
infinite sequence of states and events where $\forall i \cdot (s_i,s_{i+1})
\in T \wedge a_i \in \events(s_i,s_{i+1})$ and $s_0 \in I$.

Given a typed LKS, SE-LTL formulas are defined by the following grammar, where
$p$ ranges over $P$, $a$ over $\Sigma$, and $t$ over $\types$:
\begin{displaymath}
  \phi ::= p \mid a \mid t \mid \top \mid \neg \phi \mid \phi \wedge \phi \mid \after \phi \mid \always \phi \mid \eventually \phi \mid \phi \until \phi
\end{displaymath}
Given a path $\pi$, the semantics of a formula is the standard one of LTL with the addition that $\pi \models a$ iff $a$ is the first event of $\pi$ and $\pi \models t$ iff $a$ is the first event of $\pi$ and $\typeof(a) = t$.
$M \models \phi$ means that $\phi$ holds in the typed LKS $M$, that is, for every path $\pi$ of $M$ we have $\pi \models \phi$. Given a formula $\phi$ the goal of a model checker is to find a path $\pi$ such that $\pi \not\models \phi$. We will denote the first such counter-example, if it exists, by $M(\phi)$.
%
Given a path $\pi$, $\encode{\pi}_i$ is a formula that exactly characterises the prefix of $\pi$ up to $i$, defined as $(\encode{s_0} \wedge a_0) \wedge \after (\encode{s_1} \wedge a_1) \wedge \ldots \wedge {\after}^{i-1} (\encode{s_{i-1}} \wedge a_{i-1})$,
where ${\after}^i$ is a nesting of $i$ ``next'' operators
and  $\encode{s}$ is a formula that fixes the values of the propositions of state $s$, defined as the conjunction of all propositions appearing in $\labels(s)$ and all negated propositions in $P - \labels(s)$.

Following~\cite{scenarioexploration}, our interactive exploration technique is specified by a set of scenario exploration operations. The state of the exploration is a tuple $(\phi,\pi,i,\Phi)$ where $\phi$ is the formula being model checked\footnote{To simplify, in the remaining of the paper we will present the technique in the context of counter-example exploration in model checking, but it can obviously be also used for exploring the valid behaviours of a system with additional arbitrary constraints specified over it, by just running the model checker on the negation of their conjunction and interpreting the resulting set of counter-examples as witnesses of the system's behaviour.}, $\pi$ the current counter-example on display, $i$ the state the user is focused in, and $\Phi$ a function mapping each path index to a formula that characterises the set of states and transitions the model checker is allowed to explore at that point. Notation $\Phi \oplus \{i \twodots j\} \mapsto \psi$ will denote an update on this last function, that maps every index between $i$ and $j$ to $\psi$, keeping all other indexes intact. When updating a single index $i$, the notation will be simplified to $\Phi \oplus i \mapsto \psi$.

When first checking a property $\phi$ this state is initialised as $(\phi,M(\phi),0,\Nat \mapsto \top)$.
Basic navigation operations can then be used to inspect the counter-example, namely $\forward(\phi,\pi,i,\Phi) = (\phi,\pi,i+1,\Phi)$ and $\backward(\phi,\pi,i,\Phi) = (\phi,\pi,i-1,\Phi)$ (for $i>0$).
At any point $i$ it is possible to ask for a new counter-example that differs only in the outcome of the previous event, a useful operation to explore non-determinism. This operation is defined as $\nextstate(\phi,\pi,i,\Phi) = (\phi,M(\varphi), i, \Phi \oplus (i \mapsto \Phi(i) \wedge \neg \encode{s_{i}}) \oplus (\{i+1 \twodots\} \mapsto \top)) $, where $\varphi$ is $ \phi \vee \neg (\encode{\pi}_i \wedge {\after}^{i} (\Phi(i) \wedge \neg \encode{s_{i}}))$.
By repeatedly applying $\nextstate$ all possible outcomes of the previous action will eventually be enumerated (or possible initial states when $i=0$). Notice how $\Phi$ is used to trim a branch of the behaviour tree when this operation is selected, but maintains memory of previously trimmed branches while inspecting a trace with $\forward$ and $\backward$.

Similarly, it is possible to ask for a new counter-example that picks a different next event of the same type. This operation is defined as $\nextevent(\phi,\pi,i,\Phi) = (\phi,M(\varphi), i, \Phi \oplus (i \mapsto \Phi(i) \wedge \neg (\encode{s_i} \wedge a_{i})) \oplus (\{i+1 \twodots\} \mapsto \top) $, where $\varphi$ is $\phi \vee \neg (\encode{\pi}_i \wedge {\after}^{i} (\Phi(i) \wedge \encode{s_i} \wedge \neg a_i \wedge \typeof(a_i)))$. Notice how $\Phi$ keeps track that the branch starting in $\encode{s_i}$ and labeled with $a_i$ has already been explored. 
To ask for a new counter-example with a specific type $t$ for the next event, operation $\newevent_t(\phi,\pi,i,\Phi) = (\phi, M(\varphi), i,\Phi \oplus \{i+1 \twodots\} \mapsto \top)$ can be used, where $\varphi$ is defined as $\phi \vee \neg (\encode{\pi}_i \wedge {\after}^{i} (\encode{s_i} \wedge t))$.



\section{Implementation}
\label{sec:implementation}

Electrum is an extension of the popular Alloy formal specification
language, developed for the analysis of dynamic systems. An Alloy
model consists of a set of static signatures and relations (of
arbitrary arity). Properties can be specified in an extension of
first-order logic: apart from the standard connectives and
quantifiers, Alloy supports closures and some derived relational logic
connectives, such as composition (\a{.}) or Cartesian product
(\a{->}). To make the verification decidable, the user must specify a
scope setting the maximum size of all signatures. Counter-examples are
depicted graphically with user-customisable themes. In Electrum,
signatures and relations can be declared mutable (with keyword
\a{var}) and properties can be specified using linear temporal logic
connectives (including past ones) and primed expressions (denoting
their value in the next state) in addition to Alloy connectives.

\begin{figure}[t]
\begin{alloyfig}[numbers = none, xleftmargin=0pt, xrightmargin=0pt]
open util/ordering[Key]
sig Key {}
sig Room {
  keys: set Key,
  var current: one keys
}
sig Guest {
  var gkeys: set Key
}
one sig Desk {
  var lastKey: Room -> lone Key,
  var occupant: Room -> Guest
}

(*\color{MidnightBlue}event*) In[g: Guest, r: Room, k: Key] (*\color{MidnightBlue}modifies*) gkeys, occupant, lastKey {
   no r.(Desk.occupant) and k = nextKey[r.(Desk.lastKey), r.keys]
   gkeys' = gkeys + g->k
   Desk.occupant' = Desk.occupant + r->g
   Desk.lastKey' = Desk.lastKey ++ r->k
}
(*\color{MidnightBlue}event*) Out[g: Guest] (*\color{MidnightBlue}modifies*) occupant { ... }
(*\color{MidnightBlue}event*) Entry[g: Guest, r: Room, k: Key] (*\color{MidnightBlue}modifies*) current { ... }
(*\color{MidnightBlue}event*) Reentry[g: Guest, r: Room, k: Key] { ... }

fun nextKey[k: Key, ks: set Key] : set Key { min[nexts[k] & ks] }

fact Init { keys in Room lone -> Key and no Guest.gkeys and ... }

assert BadSafety { always { all r: Room, g: Guest, k: Key |
   (Entry[g,r,k] or Reentry[g,r,k]) and some r.(Desk.occupant) => g in r.(Desk.occupant) } } 
check BadSafety for 3 Key, 1 Room, 2 Guest, 10 Time
\end{alloyfig}
\caption{Hotel example in Electrum with events.}
  \label{fig:hotel-actions}
\end{figure}

Recently, we added the notion of event to
Electrum~\cite{electrumactions}. Figure~\ref{fig:hotel-actions} presents an
example of an Electrum model with events based on a classic Alloy example that
specifies a protocol for disposable room key-cards in a hotel.
There are 4 events in this model (check-\a{In}, check-\a{Out}, \a{Entry}, and
\a{Reentry}), each specified declaratively with relational logic and primed
expressions. The keyword \a{modifies} is used to fix the frame. The desired
safety property is that only guests registered as occupants of a room can
indeed enter that room. Unfortunately, that is not the case and the \a{check BadSafety}
command yields a counter-example trace where a guest checks in,
enters the room after checking out, a second guest checks in, and the first
guest reenters the room afterwards. This is possible because the door lock has
not yet been recoded with the new key issued by the front desk. The previous version of the Electrum
Analyzer~\cite{electrumanalyzer} already allowed the user to ask for
full alternative counter-example traces, but each one could only be inspected
independently (by navigating backward and forward in the states), making it
difficult to understand the relationship between the different
counter-examples.

\begin{figure}[t]
  \centering
  \includegraphics[width=\linewidth]{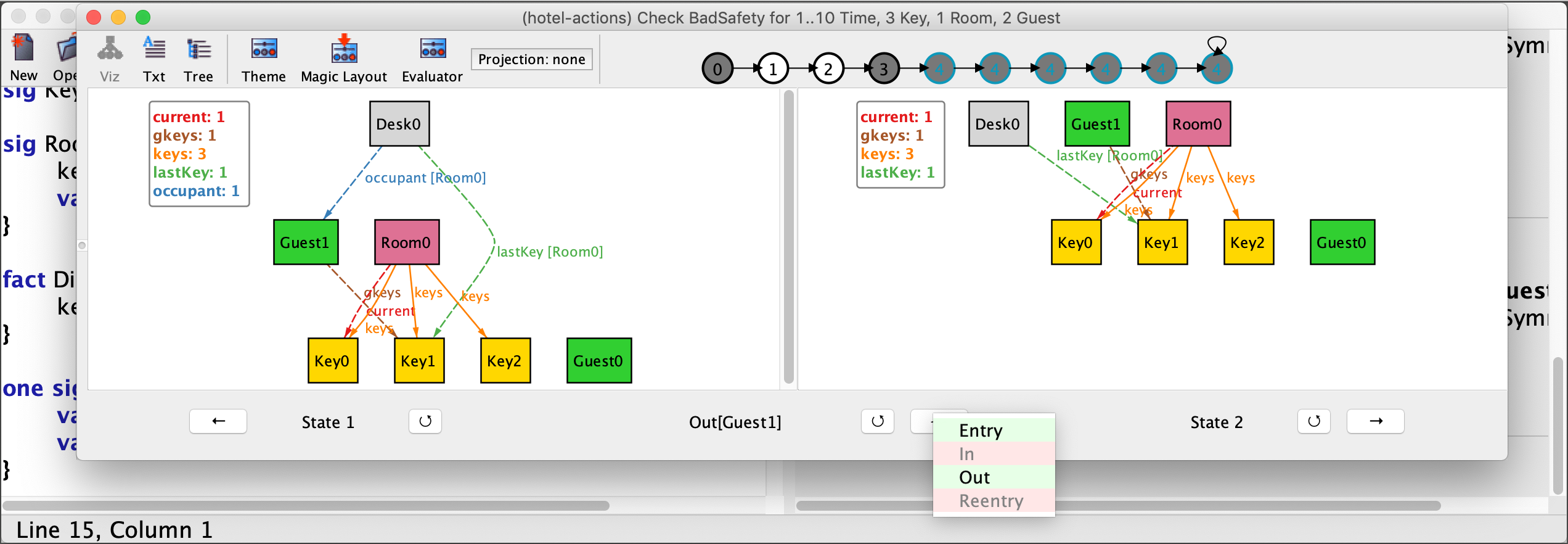}
  \caption{Exploration interface.}
  \label{fig:interface}
\end{figure}

The new prototype interface for simulation and counter-example exploration is
depicted in Fig.~\ref{fig:interface}, which illustrates precisely the
exploration of the above counter-example at $i=1$. As in the previous
version, the user can focus on a particular state of a trace by navigating
backward ($\backward$) and forward ($\forward$), using the left- and right-arrows in the bottom toolbar. However, two states are now
shown side-by-side, allowing the user to better understand what is the effect
of an event. In the top toolbar we also depict the trace and which transition
is being inspected, and the bottom toolbar in the middle shows the event that triggers the current transition. Following the formalisation in the previous section, the
user can choose a different pre- or post-state (the small ``reload'' buttons under the left- and right-panes, corresponding to operation $\nextstate$), an
event of the same type with different parameters (the ``reload'' button next to the event name, corresponding to the $\nextevent$ operation), or
a different event type to execute (the selection button in the bottom toolbar, implementing the $\newevent$ operation). When the user focuses on a
state, operation $\newevent$ is dry run on-the-fly to determine which event
types are enabled, so that when the event selection button is pressed only the
enabled events can be selected (shown with a green background, as opposed to
red for the disabled ones). In Fig.~\ref{fig:interface} we can see that after
check-in the only options are for the first guest to check out or enter the
room. Unlike in the previous version of the Analyzer, it is now easy to
understand that, for the given scope, the check-in of the second guest must
necessarily be followed by an entry or reentry of the first guest, and there
are no other possibilities to breach safety.

To assess the efficiency of the proposed technique, we measured the required time to determine which $\newevent$
operations are enabled in the different states of the first counter-example
returned by the Analyzer. Table~\ref{tab:performance} shows the results of this preliminary evaluation for
different scopes. The first column ($C$) shows the configuration (number of guests
and a list with the number of keys per room), the second the time (in seconds) to compute
the first counter-example ($T$), and then, for each state $i$, the total time (in seconds) to
compute which event types are enabled ($T_i$), and the set of enabled events ($a_i$), with the subscript
of each event type identifying the guest involved, and also highlighting the event chosen to be executed in bold. The evaluation
was performed with the bounded model checking engine of Electrum (with the
Glucose SAT solver), with maximum trace length of 10, in a commodity laptop
with a 2.3 GHz Intel Core i5 and 16 GB of RAM. As can be seen, only for $i=0$
in the last configuration did the solving of all $\newevent$ events take more
than 2s, and in most cases it is in the order of a few hundred ms. Since a
user typically needs some time to understand a state after focusing, this
delay is almost always unnoticed. Also, times tend to decrease as the user
advances in the trace: this is to be expected, since a bigger prefix of the
trace is fixed, resulting in a smaller search space for the verification
engine.

\begin{table}[t]
  \centering
  \footnotesize
  \begin{tabular}{|c|c||c|c|c|c|c|c|c|c|c|c|c|c|c}
    $C$ & $T$ & $T_0$ & $a_0$ & $T_1$ & $a_1$ & $T_2$ & $a_2$ & $T_3$ & $a_3$ & $T_4$ & $a_4$ & $T_5$ & $a_5$ & $\cdots$ 
\\
    \hline
    2[3]      & 0.07  & 0.33  & $\mathbf{I_1}$  & 0.20  & $\mathbf{O_1}$E   & 0.18  & $\mathbf{I_0}$E & 0.22 & $\mathbf{E_1}$   & 0.09 & $\mathbf{R_1}$OE   & 0.09 & $\mathbf{R_1}$OE   & $\cdots$\\ \hline
    2[1,3]    & 0.06  & 0.49  & $\mathbf{I_1}$  & 0.30  & $\mathbf{E_1}$O   & 0.27  & $\mathbf{O_1}$R & 0.23 & $\mathbf{I_0}$R  & 0.26 & $\mathbf{R_1}$     & 0.11 & $\mathbf{R_1}$OE   & $\cdots$\\ \hline
    3[2,3]    & 0.11  & 0.75  & $\mathbf{I_2}$  & 0.34  & $\mathbf{O_2}$IE  & 0.39  & $\mathbf{I_1}$E & 0.35 & $\mathbf{E_2}$I  & 0.06 & $\mathbf{R_2}$IOE  & 0.07 & $\mathbf{R_2}$IOE  & $\cdots$\\ \hline
    3[1,1,4]  & 0.58  & 1.24  & $\mathbf{I_2}$  & 0.77  & $\mathbf{O_2}$E   & 0.62  & $\mathbf{I_1}$E & 0.53 & $\mathbf{E_2}$O  & 0.23 & $\mathbf{R_2}$OE   & 0.20 & $\mathbf{R_2}$OE   & $\cdots$\\ \hline
    4[1,1,6]
        & 1.74  & 2.30  & $\mathbf{I_3}$  & 1.41  & $\mathbf{O_3}$E   & 1.10  & $\mathbf{I_2}$E & 0.94 & $\mathbf{E_3}$O  & 0.39 & $\mathbf{R_3}$OE   & 0.33 & $\mathbf{R_3}$OE   & $\cdots$\\ \hline
  \end{tabular}
    \caption{Performance of the event type enumeration (times in seconds).}

  \label{tab:performance}
\end{table}


\section{Conclusion}
\label{sec:conclusion}

This paper presented a simple technique that allows the user
  to explore the behaviours that satisfy (or falsify) an arbitrary temporal logic
  specification, with an interactive process akin to simulation. A prototype was implemented in the Electrum
Analyzer, and a preliminary evaluation showed its viability in terms of
efficiency. In
the future we intend to further
improve
efficiency by checking which events are enabled in parallel.
To show the generality of the technique we intend to apply it to other
model checkers, namely develop a counter-example 
exploration tool for SMV. Finally, we also plan to conduct a more detailed
evaluation, 
 focusing not only on efficiency, but
also on its effectiveness, namely in helping the user identify truly
different counter-examples.

\section*{Acknowledgements}

This work is financed by the ERDF - European Regional Development Fund
- through the Operational Programme for Competitiveness and
Internationalisation - COMPETE 2020 - and by National Funds through
the Portuguese funding agency, FCT - Fundação para a Ciência e a
Tecnologia, within project POCI-01-0145-FEDER-016826, and the French
Research Agency project FORMEDICIS ANR-16-CE25-0007. The third author
was also supported by the FCT sabbatical grant with reference
SFRH/BSAB/143106/2018.

\bibliographystyle{eptcs}
\bibliography{biblio}
\end{document}